\documentclass[12pt]{article}
\usepackage{epsfig}
\usepackage{palatino}
\def\beq{\begin{equation}}
\def\eeq{\end{equation}}
\def\bea{\begin{eqnarray}}
\def\eea{\end{eqnarray}}
\def\bq{\begin{quote}}
\def\eq{\end{quote}}

\def\bq{\begin{quote}}
\def\eq{\end{quote}}

\begin{document} 
\baselineskip 18pt 
\vspace*{-1in} 
\renewcommand{\thefootnote}{\fnsymbol{footnote}} 
\begin{flushright} 
hep-ph/0703060\\
TIFR/TH/07-05\\
MTA-PHYS-0702\\
\end{flushright} 
\vskip 65pt 
\begin{center} 
{\Large \bf Tevatron constraint on the Kaluza-Klein gluon of the Bulk 
Randall-Sundrum model}\\
\vspace{8mm} 
{\large\bf M. Guchait$^{(1)}$\footnote{guchait@tifr.res.in}, 
F. Mahmoudi$^{(2)}$\footnote{nmahmoudi@mta.ca} and
         K.~Sridhar$^{(3)}$\footnote{sridhar@theory.tifr.res.in}
}\\ 
\vspace{10pt} 
\end{center}
{\it 1. Department of High Energy Physics, 
                     Tata Institute of Fundamental Research,  
                     Homi Bhabha Road, 
                     Bombay 400 005, India.\\ } 
{\it 2. Department of Physics, 
Mount Allison University,
Sackville, New Brunswick,
Canada E4L 1E6.\\} 
{\it 3. Department of Theoretical Physics, 
                     Tata Institute of Fundamental Research,  
                     Homi Bhabha Road, 
                     Bombay 400 005, India.\\ } 

\normalsize
 
\vspace{20pt} 
\begin{center} 
{\bf ABSTRACT} 
\end{center} 

\noindent The Bulk Randall-Sundrum model, where all Standard Model 
particles except the Higgs are free to propagate in the bulk, 
predicts the existence of Kaluza-Klein (KK) modes of the gluon 
with a large branching into top-antitop pairs. We study 
the production of the lowest KK gluon mode at the Tevatron
energy and use the data on the top cross-section from the Run II of
Tevatron to put a bound on the mass of the KK gluon. The resulting
bound of 800 GeV, while being much smaller than the constraints
obtained on the KK gluon mass from flavour-changing neutral currents,
is the first, direct collider bound which is independent of the
specificities of the model.

\vskip12pt 
\noindent 
\setcounter{footnote}{0} 
\renewcommand{\thefootnote}{\arabic{footnote}} 
 
\vfill 
\clearpage 

\setcounter{page}{1} 
\pagestyle{plain}
\baselineskip 19pt 


\noindent The past decade has been witness to a phase of intense theoretical
activity in the area of extra space-dimensions and the resurgence
of interest in the physics of extra dimensions, originally due
to Kaluza and Klein, is due to the
new paradigm of brane-worlds \cite{brane}. For high energy physics this 
is exciting because it provides fresh perspectives to the 
solution of the hierarchy problem and also suggests the discovery of new 
physics at TeV-scale colliders \cite{dimo, anto}. 

In an attempt to find a genuine solution to the hierarchy problem
Randall and Sundrum discovered a model now known in the literature
as the Randall-Sundrum model or the RS model \footnote{More precisely,
these authors proposed two models at more or less the same time with
different features of quantum gravity in each of these. These are
now referred to as the RS1 \cite{rs1} and RS2 \cite{rs2} models. 
In our work, we will describe
and work with the RS1 model and refer to it throughout as the RS model.} 
\cite{rs1}. In the RS model, one starts with a slice of anti-de Sitter 
spacetime in five dimensions (AdS5)
with the fifth dimension $\phi$ compactified on a ${\bf
S}^1/{\bf Z^2}$ orbifold with a radius $R_c$ such that $R_c^{-1}$ 
is somewhat smaller than $M_P$, the Planck length. 
Two D3-branes called the Planck brane and the TeV brane
are located at $\phi=0,\ \pi$, the orbifold fixed points, and 
the Standard Model (SM) fields are localised on the TeV brane.
With a five-dimensional metric of the form
\begin{equation}
ds^2 = e^{-{\cal K}R_c\phi}\eta_{\mu\nu}dx^{\mu}dx^{\nu}~+~R_c^2d\phi^2 .
\end{equation} 
the model provides a novel solution to the hierarchy problem.
Here ${\cal K}$ is a mass scale related to the curvature. 
The warp factor acts as a conformal factor for the fields localised on the 
brane and mass factors get rescaled by this factor.
So $M_P=10^{19}$ GeV for the Planck brane at $\phi=0$ gets rescaled to 
$M_P {\rm exp}(-{\cal K} R_c\pi)$ for the TeV brane at $\phi=\pi$. 
The warp factor generates $\frac{M_P}{M_{EW}} \sim 10^{15}$ 
by an exponent of order 30 and solves the hierarchy problem. 
In order to solve the dynamical problem of stabilising $R_c$ against 
quantum fluctuations a scalar field in the bulk \cite{gold} with
a stabilising potential is introduced. Interesting collider
phenomenology of the model results due to the prediction of
the existence of Kaluza-Klein (KK) excitations of the graviton \cite{pheno}. 

Insights gleaned from studying the RS model \cite{holography} using the 
AdS/CFT correspondence \cite{maldacena} have suggested deformations
of the original scenario. The AdS/CFT correspondence informs us that the 
RS model is dual to a 4-d effective theory incorporating gravity and a 
strongly-coupled sector. The dual theory is conformally invariant from 
the Planck scale down to the TeV scale and it is the existence of the 
TeV-brane that breaks conformal symmetry at the infrared scales.
The KK excitations as well as the fields localised on the TeV brane are
TeV-scale composites. In effect, the original RS
theory is dual to a theory of TeV-scale compositeness of the entire SM. 
Given the unviability of such a scenario in the face of existing
experimental information, the simplest possibility is to modify the
model so that only the Higgs field is localised on the TeV brane
while the rest of the SM fields are in the bulk \cite{pomarol}.

In order to veer towards specific model realisations of such a deformation
of the RS model, flavour hierarchy, consistency with electroweak precision
tests and avoidance of flavour-changing neutral currents can be used 
as guiding principles \cite{models}. 
The location of the fermions in the bulk, or equivalently
the shape of the profiles of the SM fermions, is determined by
the fact that to get a large Yukawa coupling i.e. overlap with the Higgs
one needs to localise the fermion close to the TeV brane. Conversely, 
the fermions close to the Planck brane will have small Yukawa couplings.
The top sector needs special attention, however: the large Yukawa of
the top demands proximity to the TeV brane. However, the left-handed
electroweak doublet, $(t, b)_L$, cannot be close to the TeV brane
because that induces non-universal couplings of the $b_L$ to the
Z constrained by $Z \rightarrow b \bar b$. So the doublet needs to be  
as far away from the TeV brane as allowed by $R_b$ whereas the $t_R$
needs to be localised close to the TeV brane to account for the
large Yukawa of the top. We stress that this is one model realisation;
a different profile results, for example, in models that invokes 
a custodial symmetry or other discrete symmetries\cite{alternates}. 
It has been found that in order to avoid huge effects of flavour-changing
neutral currents (FCNCs) and to be consistent with precision tests of 
the electroweak sector, the masses of the KK modes of the gauge
bosons have to be strongly constrained. The resulting 
bounds on the masses of the
KK gauge bosons are found to be in the region of 2-3 TeV \cite{models} 
though this bound
can be relaxed by enforcing additional symmetries. A review of the
literature on this subject can be found in Ref.~\cite{review}.

There have been several studies of the phenomenology associated with
this scenario presented in the recent literature \cite{studies}.
In particular, some of these studies have focussed on graviton
production in the context of these models \cite{graviton}.
One of the interesting signals for this scenario is the production
of KK gluons. The KK gluon couples strongly to the $t_R$, with a
strength which is enhanced by a factor $\xi$ compared to the QCD coupling
where $\xi\equiv \sqrt{{\rm log} (M_{pl}/{\rm TeV})} \sim 5$. 
Consequently, it decays predominantly to tops if produced. To the left-handed
third-generation quarks, the KK gluon couples with the same
strength as the QCD coupling whereas to the light quarks its couplings 
are suppressed by a factor $1/\xi$. The problem in producing the KK
gluon at a collider, however, is that its coupling to the ordinary gluon 
vanishes because of the orthogonality of the profiles of these particles.
The KK gluon can, therefore, be produced by annihilation of light
quarks and this production mechanism has been studied in the
context of the LHC \cite{kkgluon}.

In this paper, we investigate the production of KK gluons at the Tevatron and
its decay into top pairs and use the measured top cross-section at
the Tevatron to obtain a lower bound on the mass of the KK gluon. 
Given that the Tevatron reach is limited kinematically, it is not
going to probe the range that is probed by precision electroweak
tests, FCNCs,  or by the LHC. Nevertheless, it is useful to determine what
is the {\it direct, model-independent} bound that existing collider
data can provide. To express it differently, even if a specific model
avoids the problem of FCNCs through the incorporation of a new symmetry
or by a novel choice of profiles of SM particles in the bulk
and allows for the gluon KK modes to be much lower in mass, the Tevatron
bound will still be applicable.

The cross-section for the production of a KK gluon of mass $M_*$ 
via quark-antiquark annihilation is given by
\begin{equation}
\sigma = {4\pi \over 9} {\Lambda_{q}^2 \over M_{*}^2} \int dy \sum_q 
x_1 q(x_1, M_{*}^2) x_2 \bar q(x_2, M_{*}^2) + (x_1 \leftrightarrow x_2)
\end{equation}

$\Lambda_q$ is the coupling of the KK gluon to light quarks and is equal
to $\sqrt{4 \pi \alpha_s}/5$. A KK gluon with a mass just a little above
the $t \bar t$ threshold has a very large branching into top pairs: the
branching ratio is about 92.5\% \cite{kkgluon}.

One can actually include the effect of QCD corrections quite trivially.
Since the SM gluon does not couple to the KK gluon, the gluon
in the NLO-QCD diagrams couple only to the quark legs. This is exactly 
like the Drell-Yan production of lepton pairs and so one can simply put 
in the Drell-Yan K-factor here as an overall factor to account for the
QCD corrections. This is given by
\begin{equation}
K=1+{8\pi \over 9} \alpha_s (M_{*}^2)
\end{equation}

\begin{figure}[h] 
 \centering    
\includegraphics[scale=0.5, angle=-90]{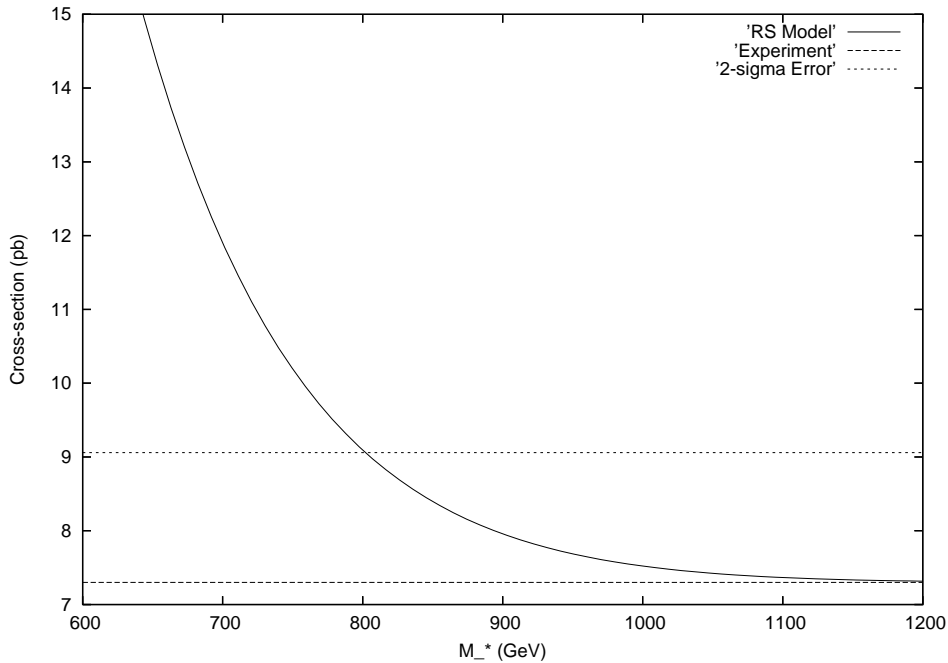}     
\caption{\it The cross-section for KK gluon production as a function of its
mass. The horizontal lines show the CDF central value for the
cross-section and the 2-sigma upper limit.}  \protect\label{fig1}
\end{figure}

In Fig~1, we have plotted the cross-section as a function of the scale $M_*$
for $p \bar p$ collisions at the Tevatron energy of $\sqrt{s}=1.96$~TeV. 
We have used the CTEQ4M densities \cite{cteq} and the parton distributions
are taken from PDFLIB \cite{pdflib}. For the QCD scale, we use $Q=M_*/2$.
We have folded the calculated cross-section with the branching ratio
of the KK gluon into $t \bar t$ which is 92.5\%.
For the experimental value of the cross-section we have used the value
presented by the CDF collaboration (averaged over all channels)
from the Run II of the Tevatron given in Ref.~\cite{tevatron} which is
quoted as $\sigma_{t\bar t} = 7.3 \pm 0.5 (stat) \pm 0.6 (syst) \pm 0.4 (lum)$.
The central value and the $2\sigma$ band
of this cross-section (with the errors added in quadrature) are 
also shown in Fig~1. We see that 
a bound of about 800~GeV results at the 95\% confidence level. 
For other choices of scale and parton distributions, the cross-section
varies by about 25\% resulting in about a 20-30 GeV in the value of
the bound on $M_*$.

In conclusion, one of the striking predictions of the bulk RS model is
the existence of KK gluons which decay into a $t \bar t$ pair. We have
computed the production cross-section of these particles at the Tevatron
and compared it with the $t \bar t$ cross-section measurement from the
Run II of the Tevatron. The lower bound on the KK gluon mass is obtained
to be about 800 GeV at 95 \% C.L. This is the first direct collider bound
on the mass of the KK gluon in this model.

 
\end{document}